\documentclass[conference]{IEEEtran}
\IEEEoverridecommandlockouts
\usepackage{cite}
\usepackage{amsmath,amssymb,amsfonts}
\usepackage{algorithmic}
\usepackage{graphicx}
\usepackage{textcomp}
\usepackage{comment}
\usepackage{hyperref}
\usepackage{layouts}

\usepackage{xcolor}
\usepackage{colortbl}
\def\BibTeX{{\rm B\kern-.05em{\sc i\kern-.025em b}\kern-.08em
    T\kern-.1667em\lower.7ex\hbox{E}\kern-.125emX}}

\newcommand{\gf}[1]{{\bf Goerschwin: #1}}
\renewcommand{\gf}[1]{}

\definecolor{dunkel}{HTML}{77BEFF}
\definecolor{hell}{HTML}{A7EEFF}

\begin{document}

\title{YAPS -- Your Open Examination System for Activating and emPowering Students\thanks{We would like to thank Gianluca Martino, Swantje Plambeck, Lutz Schammer for the support while creating, testing, and using YAPS.}}

\author{\IEEEauthorblockN{Fin Hendrik Bahnsen \hspace{2cm} Goerschwin Fey}

\IEEEauthorblockA{
\textit{Hamburg University of Technology}, 21073 Hamburg, Germany \\
\{fin.bahnsen,goerschwin.fey\}@tuhh.de}
}

\maketitle

\begin{abstract}
There are numerous e-assessment systems devoted to specific domains under diverse license models. Cost, extensibility, and maintainability are relevant issues for an institution. Ease of use and inclusion into courses are educator's main concerns. For students the user experience and fast transparent feedback plus ``better'' tests are most important.
Many exams still focus on testing memorized knowledge, instead of improving and testing skills with competence-oriented learning support and examinations, respectively.

We discuss design decisions and present the resulting architecture of YAPS -- Your open Assessment system for emPowering Students.
YAPS has been used for very diverse lectures in logistics, computer engineering, and algorithms for exams, but also for empowering students by fast feedback during the learning period.
We report on results in a basic lecture on Computer Science for Mechanical Engineers.
\end{abstract}


\section{Introduction}

\gf{Who is missing in thanks?}

E-assessment in university education has a number of advantages if used carefully.
The main advantages for students should be more competence-oriented testing. Written exams in many domains have to restrict to simplified mechanical problem solving instead of competence-oriented testing. A main reason is the effort needed to evaluate solutions for more extensive and flexible exercises. Another reason is the lack of feedback during the assessment. For lecturers the advantages may be in more efficient evaluation of tests and exams, as well as in lowering the barrier to get informed feedback about student's level of expertise during the learning period. Finally, on the institutional side lowered cost for examinations and also an improved experience for students are driving forces.

In \cite{HMB:2009} so-called blended learning combining team experience, practice, theory, and reflection has shown to be activating students and to significantly improve pass rates. Personal involvement into practical exercises, individual learning, and feedback are three of the key aspects.

A web-based architecture has been proposed in~\cite{Jun:12} and has similar requirements as we do, but has taken some different design decisions in technical solutions and focused on the architectural design of the system with simplicity as one of the main drivers. We consider a broad set of requirements and discuss how these requirements guide our system design.

MapleTA~\cite{Map:21} is a strong commercial tool with underlying direct connection to Maple with all its mathematical libraries and models. Moreover, MapleTA allows to create custom-made exercises by linking self-made plugins into the exam.
Evaexam~\cite{Eva:21} is another commercial tool that even allows for a mix of paper-based and online exams. While the configurability with respect to exercise types is reduced, the full process including the institutional setup and splitting of responsibilities is built into the system. However, for commercial platforms license cost, dependence on a single supplier, and restricted options for adaptability due to a closed source model make the adoption difficult. Moreover, intellectual property in terms of exercises and custom extensions are tightly bound to such a system. Future migrations to the cloud may move these assets from the examiner and the examining institution closer to the supplier of the examination system.

\gf{What else?}

In this paper we describe the web-based YAPS -- Your open Assessment system for emPowering Students\footnote{You may decide whether YAPS is just an acronym or the pronunciation should be close to some sound one might make while solving an exam as participant, preparing an exam as a lecturer, or looking at the results as an examiner. Using YAPS, the examiner is likely to utter a mix of ``YES'', ``PASS'', and ``SUCCESS''!} On the user side, YAPS has been designed to support students in their learning progress and during exams with a strong focus on competences instead of knowledge only. 
On the backed side, extensibility, usability, and maintainability were the drivers.
We take a technical perspective in explaining YAPS and report from our experiences with the system.

Our specific contributions made with YAPS are
\begin{itemize}
\vspace{-1ex}
\setlength\itemsep{0em}
\item student support as a main driving concern,
\item institution wide usage, separation, security, and authorization,
\item intuitive usage and reporting,
\item based on most recent web technologies and virtualization.
\vspace{-1ex}
\end{itemize}

\gf{We demonstrate how our system meets our pre-defined requirements and we provide empirical data to support our claims}

This paper is structured as follows. In Section~\ref{sec:req} we state the main requirements and features that drive system development. Section~\ref{sec:workflow} explains how YAPS facilitates ease-of-use and adaptability from an examiners's point of view and how institution-wide management is supported. The resulting system architecture is explained down to a technical level in Section~\ref{sec:architecture}. We sketch exemplary exercises to give an intuition of the student's view and benefits in Section~\ref{sec:example}. We show compliance with requirements and present results based on student's feedback through surveys in Section~\ref{sec:results}. Section~\ref{sec:conclusions} concludes.

\section{Requirements and Features}
\label{sec:req} 

We describe the most important requirements for YAPS in the following. The list is not comprehensive but points to the main goals driving the development beyond typical aspects like stability, security, etc. Some of these requirements put a direct demand for features to be offered by the examination system. 

\begin{itemize}
  \item Competence-oriented exercises 

  While some simple questions that target the assessment of knowledge shall be available, 
  the system shall provide options for competence-oriented assessment. 
  These exercises shall be practical assignments to solve pre-defined problems. Practical support that would be available in a real environment shall also be provided while solving the exercises.

  \item Usability

  From the student side the system shall be intuitively usable. Student's should not be hindered by the system's interface, but focus directly on solving the exercises. 

  Lecturers and examiners shall be able to configure the system transparently with a very good understanding of the student's view during the exam and a very good understanding of their settings.

  \item Very fast feedback

  During the exam required feedback shall be provided directly. This does not regard the grading, but feedback for problem solving. 

  After the exam the results shall be provided to the student directly after a short waiting period. However, examiners must be able to view at results before informing students.

  \item Extensibility 

  The examination system shall be usable for a wide variety of lectures from different disciplines. 
  In the general setup the type of exercises shall be freely configurable.

  There are very different use cases; besides a standard exam concurrently running self-tests are useful to provide fast feedback to students. Thus, an extension of a pre-defined exam while this is running shall be possible. 

  Randomization of exercises shall be possible to have large classes using the same exam at different points in time.

  \item Preservation

  Unexpected intermediate events of technical or non-technical origin may cause disruptions or disturbances. Thus, for individual students their progress and intermediate states shall be traceable. 

  After the examination results must be available for a legally defined amount of time. Thus, the full exam shall be reproducible.

  \item Maintainability

  On the client side hardware and machines available may change regularly. Thus, thin non-invasive portable clients shall be supported.

  The infrastructure shall be portable and maintainable from as few access points as possible.

  Maintenance of exams including custom exercises shall be easily and exclusively accessible to the respective examiners.

  \item Data privacy

  Information related to individual persons, specifically any student information, must be strictly separable from all other data. After the formally foreseen storage period this data must be completely erasable.

  Only authorized persons shall be able to access data related to individual persons.
\end{itemize}

\begin{figure}[t]
\centering
\includegraphics[width=.5\textwidth]{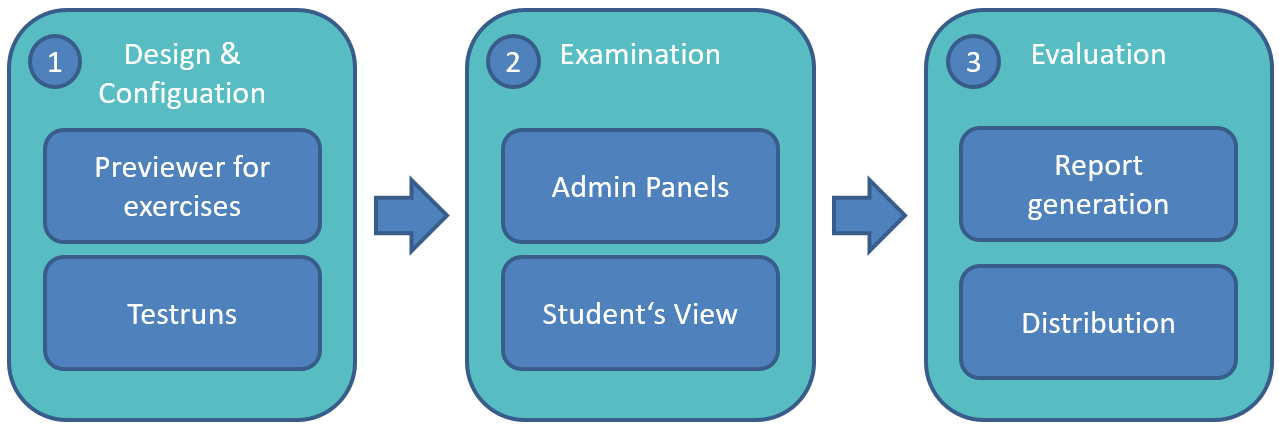}
\caption{Workflow per Exam} \label{fig:workflow}
\end{figure}
\section{Workflow}
\label{sec:workflow} 

Before explaining YAPS' architecture and internals that are designed to meet all of the above requirements, we sketch the workflow for each individual exam.
\gf{Currently this is the workflow per exam, We could add a subsection for the institutional work flow}
Figure~\ref{fig:workflow} provides an overview. An exam starts with (1)~a design and configuration phase~(1), proceeds to (2)~the actual examination, and finishes with (3)~evaluation and reporting. 

\subsubsection{Design and Configuration}
During design and configuration the individual exercises are put together. Also custom exercises may be created which may require some programming to embed these custom blocks into the overall system. 
A previewer for exams helps to design exercises interactively in a ``What you see is what you get'' manner. All of the exercises, how these would be displayed to the students, and what can be entered are available in this previewer. When adapting any piece in the configuration the previewer directly reflects such changes.

Besides configuring the exercises, also any administrative setup, e.g., when an exam starts, which students are entitled to join the exam is done at this stage. All of these settings are based on human readable information in files, so the data is easily accessible.

Trial runs of the real exam can then be performed. These trial runs are directly initiated by the responsible examiner on demand. This is important to ensure high-quality exams, understand the consequences of the evaluation, and potentially apply and re-test required changes. These trial runs follow the same procedure as any other exam.
Moreover, the roll-out procedure for an exam is the same, authorized participants may be pre-defined to restrict access to registered examinees. 

\subsubsection{Examination}
On the participant's side the examination follows a standard flow for each participant:
\begin{enumerate}
\item Authorization via the institution's typical mechanisms and checking whether students feel ready to take the exam.
\item Explanations, instructions, and terms for the exam are provided to the student and to be accepted by the student before starting the exam.
\item Solving the exercises within a pre-defined time. Whether this time starts globally for all participants, those in a certain room, or when the student accepts the terms can be configured to support different use cases. During this time students can freely navigate between individual exercises and the state per exercise is preserved.
\item Submission of the exam by the student or upon reaching the pre-defined submission time.
\end{enumerate}

While the examination takes places, an ``admin panel'' provides access to control individual examination process, groups or all participants. This admin panel is used by all supervisors who may not have a detailed understanding of how an exam is designed. The admin panel provides multiple features. For example, students that lost their credentials can be logged in by-passing the institutions mechanisms, time extensions can be provided, individuals can be validated based on physical examination of identity cards and student cards. ``Supervisor level'' access into the admin panel is required for these tasks and is pre-configured for authorized users. Moreover, the number of active participants can globally be monitored by authorized users with ``admin level'' access.

Indeed the exam may even be extended by further exercises in this phase. This feature is specifically useful for running a growing testing instance concurrently to the standard lecture time, so students can train themselves independently.

\subsubsection{Evaluation and Reporting}
The final phase is also supported via the admin panel for users with ``admin level'' access. First, the overall evaluation including grading of all individual submissions takes place. This evaluation can run fully automated or, if required, certain parts can be set for manual evaluation. If required, additional information can be added at this stage, e.g., bonus points earned when taking part in intermediate tests. The result of the evaluation is provided by reports per student. Each student gets direct feedback for the answers provided. Expected solutions may be included in these reports. Once the evaluation has been cleared by the examiner the reports can either be directly emailed to all participants or they can be accessed via browser-based login into the examination system for downloading the report.

All examination data can then be downloaded for storage of personal data and examination data of all participants. This data is available through individual reports, spreadsheets, and -- if required -- by a full export of the underlying data base.
After deletion of the exam instance no personal data remains in any of the running systems. Only the exported data documents the full exam. The exam itself may be reproduced from the knowledge base and configuration data which is stored separately and persistently.

\section{Architecture}
\label{sec:architecture}
On the server side, the architecture relies significantly on the container software \textit{Docker} \cite{docker}. A container is a lightweight and secure way to run isolated instances of a software in a virtual runtime. In general all exams share a joint server with individual containers for each part of the YAPS bakend to guarantee isolation between exams.
For a specific lecture there is a common history and knowledge base for all exams stored in a versioning system relying on {\em git}. However, each individual exam then has its own lifetime and any personal data is decoupled from the knowledge base using so-called docker volumes.
\subsection{Single Exam}

\begin{figure}[t]
\centering
\includegraphics[width=\columnwidth]{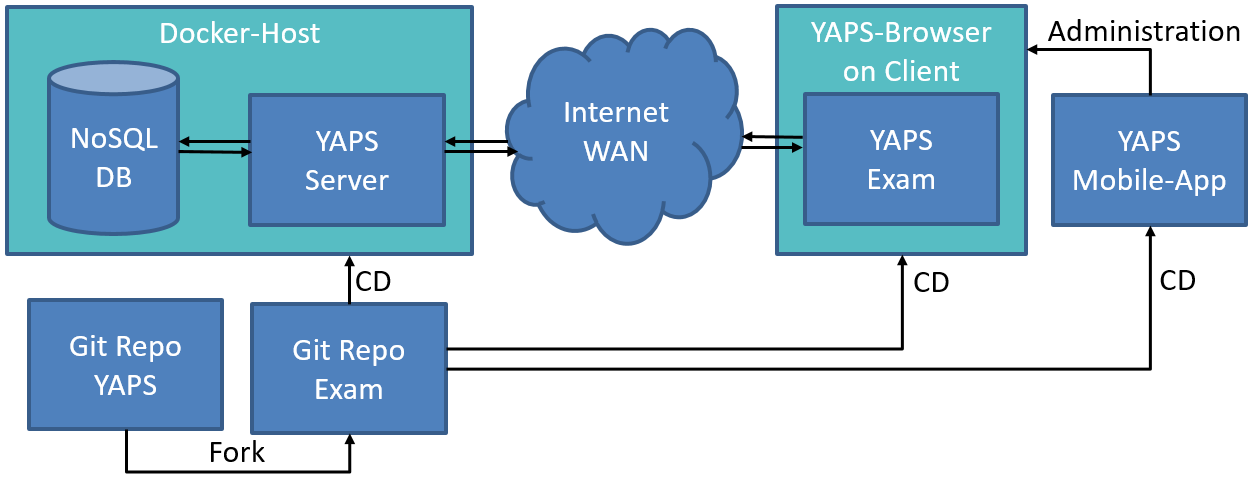}
\caption{Architecture per Exam} \label{fig:architectureExam}
\end{figure}

Each examiner works with a separate fork of the exercise framework. As shown in Figure \ref{fig:architectureExam}, this is automatically deployed on the central server infrastructure via continuous deployment (CD) using a {\em gitlab} infrastructure. The content that the examiner creates in this repository is completely independent of others. Nevertheless, all users of the system can collaboratively share and work together on content via the pull request and merge system naturally available in {\em git}.

\subsubsection{Exercise Framework}
Each exercise is represented by a stand-alone static web page. The web page already contains the code to display the exercise in the desired formatting or to reload answers from a participant and, if necessary, the results of an evaluation. Typically, the code for automatic assessment is also embedded directly into the static code of the exercise, but this is not delivered to the students' client systems at exam time. Using such an interface, exercises can be developed and also tested independently. The exercise interfaces to the system with the following methods: (1)~a method to store the participant's input, (2)~a method to store the assessment result, and (3)~a method to reload such data provided by the server from the database.

Using the designated exercise framework ensures the correct implementation of the described interface for each exercise. Nevertheless, there is the possibility to implement the interface itself and to design very individualized exercises. This gives the examiner full flexibility, as there are almost no restrictions. 

Moreover, the routines for saving and loading data are defined asynchronously. By this, even external services can be accessed without any problems. For example, the client system can simply wait until an external compile service provides results for a programming exercise. Due to the asynchrony at this point, the loading process of other exercises is not interrupted or slowed down by these waiting times.

\subsubsection{Storage of Student Data}

Data provided by different exercise types or even by individually designed exercise types are extremely heterogeneous in nature. Schema-based database models are difficult to use in this kind of application. In our case the data is stored in a document based no-SQL \textit{Mongo} database \cite{mongo}. In order to store the participant's input, the server simply stores the data as a document in {\em JavaScript Object Notation} (json) in the corresponding data set. In the load routine the server returns the document exactly the same way it was saved.

In this sense the database acts as a remote object storage without being bound to a fixed schema. As long as an exercise processes the data that it has provided itself, this works reliably and at the same time avoids the complex dependencies that are imposed by a special data schema, for example. By definition, there is no exercise that requires special server-side code.

\subsection{Components of a Single Exam}

\begin{table}
	\caption{Technologies and Libraries} \label{tab:technologies}
	\begin{center}
		\begin{tabular}{|p{.18\columnwidth}|p{.32\columnwidth}|p{.35\columnwidth}|}
			\hline
			\rowcolor{dunkel}
			\textbf{Component} & \textbf{Technology} & \textbf{Libraries} \\ 
			\hline
			Previewer   & Live Viewer & Docker-Development-Container \\
			\hline
			Exam-Server & noSQL DB & MongoDB \\
			& Webserver & Express \\
			& Languages & Typescript, NodeJS \\
			\hline
			YAPS-Server & Virtualization & Docker \\
			& Webserver/Networking & nginx \\
			& Languages & NodeJS, Typescript, Python, HTML \\
			\hline
			Client      & Framework & Chrome/Electron \\
			& Languages & NodeJS, Typescript, Polymere, SocketIO \\
			\hline
		\end{tabular}
	\end{center}
\end{table}

A deeper insight into the generic infrastructure is gained by taking a closer look at the individual components of an exam. We give a brief overview of technologies and related tools used for all of the components in Table~\ref{tab:technologies}.

\subsubsection{Exam-Server}
For each exam, a separate virtual docker context with an associated virtual network adapter is run on the YAPS-server. Each exam repository contains the description of a docker compose -- a collection of different containers that provide different services internally or externally over the network. This means that an examiner can determine via configuration in the repository which services should be provided for this specific exam.

In the simplest case, this is only a container for the Mongo database and a container for the YAPS exam server, but further services can be added. For example, a compile server that is used in programming tasks to execute code that exam participants write in an arbitrary programming language during the exam.

\subsubsection{Client and Admin Frontend}
The YAPS-server also serves as a web server and delivers the web pages for the client systems as well as for the admin panel using an \textit{Express} \cite{express} webserver on \textit{NodeJS} \cite{nodejs}. For a particularly high level of stability, both applications were developed independently of each other and rely on modern web technologies such as web components. In order to keep the cost for further development low, the \textit{Polymer} \cite{polymere} library was used for both applications. For the real-time communication between server and client the high level library \textit{SocketIO} \cite{socketio} is used.

Both, the client application and the admin panel are used generically for all exams. The individual exams adapt the generic templates of the applications according to a configuration file.

\subsubsection{Configuration}
An individual configuration file in the exam repository controls the entire process of the examination. This includes the examination process as well as the selection of the exercises. This procedure allows exercises that are not to be included in the exam to be stored in the exam repository as well. Thus, the exam repository becomes a knowledge base for future exams. By rewriting the mentioned configuration file, exercises are selected quickly from the existing pool and combined to form new exams. Furthermore, the configuration serves to control essential aspects of the exam, such as the time control or the access authorization to the exam and the admin panel.

\subsubsection{Roll-out}
Each exam repository is linked to a dedicated \textit{gitlab-runner} \cite{gitlab-runner} that allows exclusive access to the exam's docker context. Tagging a commit in the repository triggers an automatic roll-out process, referred above as continuous deployment, that starts the services described in the configuration in isolation on the server.

A higher-level structure takes over the orchestration and makes the services that are to be externally accessible via the network of the server available under dynamic urls. Using this mechanism, the examiners can adjust the infrastructure for their own exams as often as they like without any additional response of an system administrator. This is useful, e.g., for trial runs before starting the actual exam.

\subsubsection{Examination}
With the completion of the continuous deployment, the exam is immediately ready for use. The system relies on several different variants of access control. For example, when the client system is opened on a computer, it must be explicitly authorized by a supervisor before a participant can log in.

The status of the client systems is managed on the server side and synchronized regularly. As mentioned above, the admin panel is used to change the client state and to monitor the client instances.

\subsubsection{Grading and Reporting}
As already mentioned, the exercises also contain the code to evaluate the exercise completely or partially. To ensure that the evaluation is carried out in an absolutely comparable manner, a Chrome browser is running in headless mode on the server side. During the evaluation this browser is remote controlled by the \textit{Puppeteer} library \cite{puppeteer} from Google. The setup ensures that only this internal browser can load the full task code that includes the assessment functionality. At the same time the ''print to pdf'' function of the browser creates a document containing the participant's answers and the result of the assessment.

The system finally combines the PDF documents of a participant into a multi-page report and adds an overview page at the beginning of the document. This page contains an overview of the evaluation as well as bonus points, a grade chart, and the grade. The design paradigm used for exercises ensures that the assessment view of an exercises explains as transparently as possible how the assessment came about. For example, sub-scores are displayed at various places. At the same time, the report fulfills all requirements for archiving the exam results properly.

Once the assessment has been finalized, the system can automatically distribute the reports to the participants via e-mail or deliver them to a specific computer in an in-place setup. This way, exam reviews by the exam's participants can be realized easily and efficiently.

\subsection{Institution Wide View}
\begin{figure}[t]
\centering
\includegraphics[width=\columnwidth]{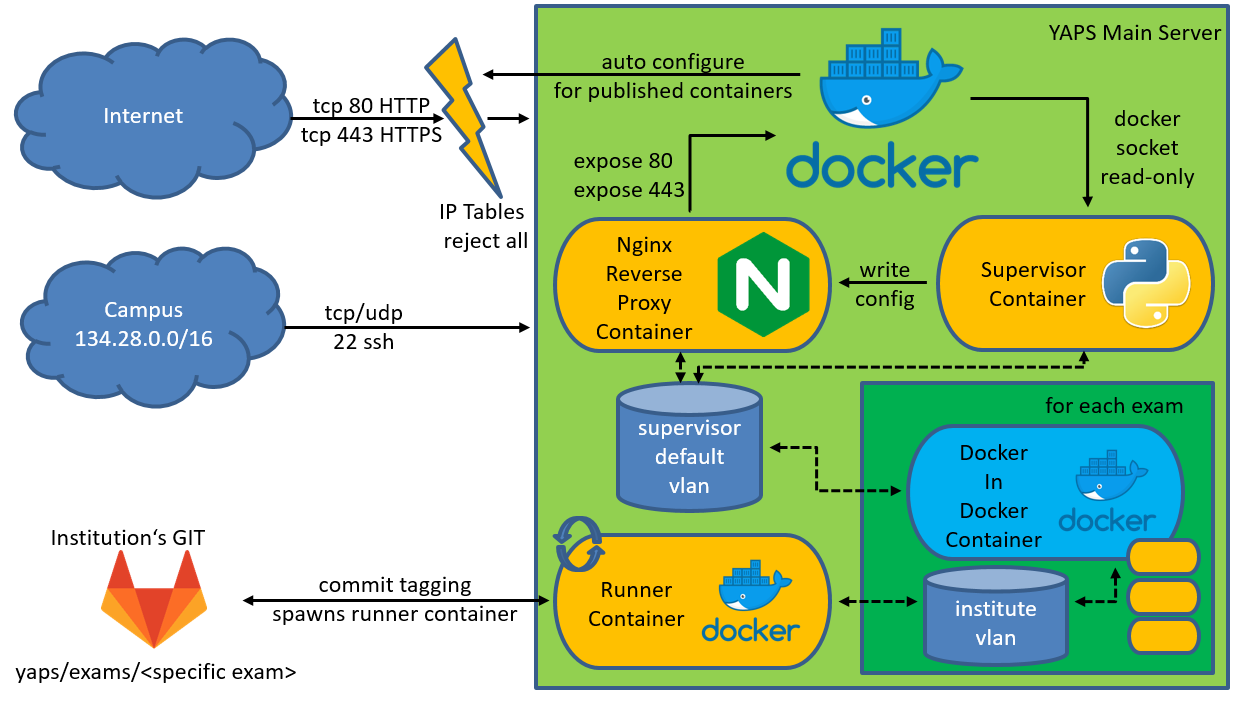}
\caption{Architecture for the Institution} \label{fig:architectureInstitution}
\end{figure}

The advantages of the architecture described so far become clear if it is viewed once again from an institutional point of view. Only a single server is required for the YAPS exam system, which keeps the administrative effort low and centralized. Due to the layout chosen for the continuous deployment, the examiners still remain very flexible in the design, e.g., by including proprietary software as a service using docker containers. Using templates in the web application for the exam client and the admin panel makes these components usable for all exams. Thus, changes in these can be applied simultaneously for all exams.

Figure \ref{fig:architectureInstitution} shows the YAPS server architecture in more detail. A container with privileged rights runs on the server to communicate with the docker service of the server. A python-based supervisor script regularly checks which of the available containers should provide a service to the external network. If a new container is found, the configuration of an \textit{nginx} \cite{nginx} proxy server is dynamically adjusted.

A {\em docker-in-docker} (dind) container is created fully automatically for the individual exams. In this virtual docker context, the services for an exam can be operated completely isolated in their own virtual sub-network. This ensures the data privacy of the sensitive participant data. At the same time, server secrets such as certificates are located at the level of the host machines operating system and cannot be spied on directly or indirectly by the examiners due to gitlab-runner giving access only to the designated docker context. \gf{Den Satz direkt hiervor verstehe ich nicht. Heisst "supervisor level" hier root level oder ist das ein Docker-Ausdruck bzw. gitlab-CD-Ausdruck?}

At the same time, integration with the institution's gitlab creates a strong opportunity for collaboration among examiners, e.g., in the development of new types of exercises or specialized solutions. Furthermore, the continuous deployment reduces the administrative effort to a minimum.

\begin{figure}[t]
	\centering
	\includegraphics[width=\columnwidth]{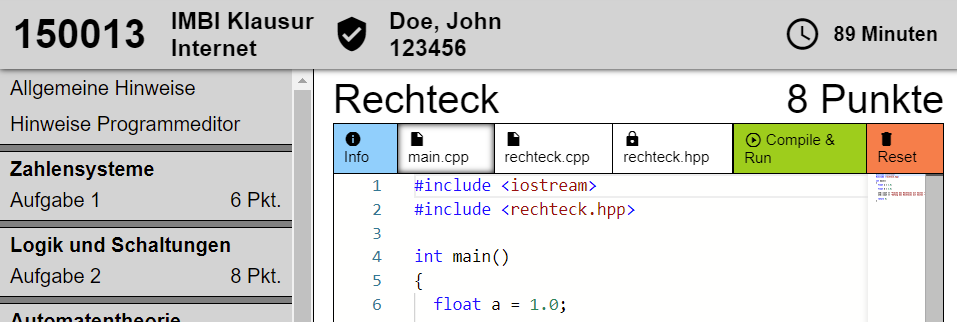}
	\caption{Client view of a programming exercise.} \label{fig:programmingExercise}
\end{figure}

\begin{table*}[t]
\centering
\caption{Compliance with requirements}
\label{tab:compliance}

\begin{tabular}{|p{.195\textwidth}|p{.78\textwidth}|} \hline
\rowcolor{dunkel}
  {\bf Requirement} & {\bf Solution} \\ \hline
\rowcolor{hell}
  Competence-oriented exercises  & Full configurability through custom programmed exercises and server-side on-demand feedback  \\ \hline
\rowcolor{hell}
  Usability & \\ \hline
  Students & Intuitive browser-based interface\\ \hline
  Examiners & Pre-defined templates for many exercises like multiple-choice, graph-modification, programming \\
              & Configuration is file-based, knowledge base per exam stored in a versioning system, programming level accessible for transparency\\ 
              & Admin panel provides all required functionality for the supervisor team during the exam\\ \hline
\rowcolor{hell}
  Very fast feedback & \\ \hline
  During the exam & Parts of evaluation can be triggered interactively, serve-side responses can be displayed, e.g., for programming exercises compiler output and output from program executions are available\\ \hline
  After the exam & Automated evaluation, reporting by email, triggered through admin panel\\ \hline
\rowcolor{hell}
  Extensibility  & \\ \hline
  Freely configurable exercises & Full access to the code-base, simple programming interface per exercise for display, interaction, and evaluation\\ \hline
  Extension while running & Data-base is extensible while running to add further exercises \\ \hline
  Randomization & Per participant randomization seeds are generated, these seed serve to generate the individual exams on the server side\\ \hline
\rowcolor{hell}
  Preservation & \\ \hline
  Traceability per student & Interactions with the server are stored in the data base, the full state of a participant's solutions after each interaction is available \\ \hline
  Reproducibility  & Versioning system allows to uniquely reproduce the full state of the server and each exam for any point in time\\ 
              & Participants data can be exported including the data base \\
              & Versioning and interaction storage facilitate full reproducibility of a full exam and each participant's progress\\ \hline
\rowcolor{hell}
  Maintainability & \\ \hline
  Thin clients & Fully browser-based access for participants and supervisors, no installation beyond browser in kiosk mode \\ \hline
  Infrastructure & Separation of components like server, client, exam configuration per lecture also separate concerns\\ 
                 & Knowledge base directly grows in the versioning system per exam, but can also be copied to the YAPS overall system, e.g., to deploy new types of custoom exercises \\ \hline
  Exams and custom exercises & File-based configuration and full access to code-base\\ \hline
\rowcolor{hell}
  Data privacy & \\ \hline
  During the exam & Protection by a dedicated data base server \\ \hline
  After the exam & Export of all personal and examination data related to participants, none of this data remains in the system \\ \hline
  Authorization & Admin panel is the only entry point for evaluation and accessing participants data; different access levels are available where only persons with ``administration level'' can access sensitive data. \\
               & Examiner decides on access levels for individuals \\ \hline
\end{tabular}
\end{table*}

\begin{figure}[t]
	\centering
	\includegraphics[width=\columnwidth]{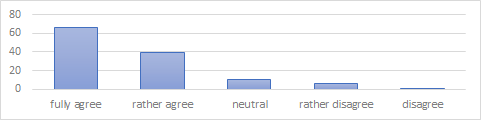}
	\caption{Answer for ``Were the exercises well-arranged?''} \label{fig:arranged}
	\centering
	\includegraphics[width=\columnwidth]{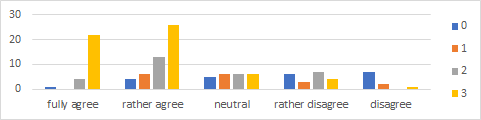}
	\caption{Answer for ``Did you feel well-prepared for the system?'' versus participation in 0, 1, 2, or 3 intermediate tests}  \label{fig:prepared}
\end{figure}
\section{Exemplary Exercise}
\label{sec:example}

Programming exercises are a specifically suitable example for competence-oriented testing in YAPS.
Figure~\ref{fig:programmingExercise} shows how programming exercises are embedded in the system's user interface (UI). Participants can switch between tasks in the navigation bar on the left, while the upper title bar displays basic participant information, such as, e.g., the name, the matriculation number, or the remaining time. The central area displays the exercise.

In this particular case, the upper tab bar of the exercise can be used to switch between different files that are part of a program. The first info tab explains the programming task that is to be solved by modifying or completing the provided code files. 

The program can be compiled at any time by the participant as a test. For each compile operation, the compile server creates a new docker container in the virtual context of the exam and communicates the results back to the exercise web page. The UI then displays compiler messages and, if necessary, program output a few seconds later. The participant can also reset the task to its initial state via a reset button. By this, participants have a natural interface for programming and get all the compiler feedback, e.g., hints for syntactic errors or simple problems in their programs that they would get in a standard programming situation in practice.

For the evaluation, the static exercise web page also contacts the compile server in order to check specified unit tests for the participant's program. The assessment is then based on which unit tests were passed. The asynchronicity of the evaluation function mentioned at the beginning plays a decisive role here.

Especially this or similar programming tasks illustrate why there must be different views for some exercises. If the view from Figure~\ref{fig:programmingExercise} were to be included unchanged in the report, only parts of the program would be visible in the report. The programming task has a special archiving view in which all files are output one after the other and finally a table with an overview of the tested test cases.

As a result participants solve a real world programming challenge with all typical support and get detailed feedback including their code in their individual reports.

\section{Results}
\label{sec:results}

First, Table~\ref{tab:compliance} shows in detail how each of the requirements set forth in Section~\ref{sec:req} is satisfied by YAPS. 

Second, several exams have been performed with YAPS already. While the focus of this paper is on the technical aspects of YAPS, we provide initial results on the user experience coming from a survey. The data was collected from 161 first semester students participating in a mandatory lecture on Computer Science for Mechanical Engineers. 
Figure~\ref{fig:arranged} shows that most participants were pleased by the arrangement of exercises in YAPS. Moreover, Figure~\ref{fig:prepared} shows that participants who took part in all three intermediate tests felt well-prepared to use the system. This is a strong indication to further expand the capacities for feedback during the learning period.

\section{Conclusions}
\label{sec:conclusions}
YAPS is an extensible online examination system. Extreme configurability facilitates usage for lectures from a wide range of domains. Most specifically, competence-oriented examination using problem solving tasks helps students to assess their practical skills. Even using YAPS for self-assessment concurrent to lectures is possible.
YAPS has already been used for computer science, computer engineering, algorithms, and logistics courses. 

Future work in YAPS will focus on adopting even more flexible approaches for grading including free text and ``almost correct'' solutions. A next step will then be to use these features in supporting students in understanding and resolving their mistakes in concurrent self-training settings.

\bibliographystyle{ieeetr}
\bibliography{examsystems}

\end{document}